\documentclass[12pt]{article}
\usepackage{amsmath,amssymb}
\usepackage{indentfirst}
\usepackage{cite}
\usepackage[utf8]{inputenc}
\newcommand\tr{\mathop{\mathrm{tr}}\nolimits}
\newcommand\Tr{\mathop{\mathrm{Tr}}\nolimits}
\newcommand{\diff}{\mathrm{d}} 
\makeatletter

\@addtoreset{equation}{section}
\makeatother
\begin{document}
	
	\begin{titlepage}
		
		\begin{center}
		
		\hfill UT-18-15\\
		
		\vskip .75in
		
		{\Large\bf
		A calculation of the gauge anomaly
		
		with the chiral overlap operator 
		}
		
		\vskip .5in
		
		{\large
		Taichi Ago
		}
		
		\vskip 0.25in
		
		{\em 
		Department of Physics, The University of Tokyo, Tokyo 113-0033, Japan}
		
		\end{center}
		
		\vskip .5in
		
		\begin{abstract}
			We investigate the property of the effective action with the chiral overlap operator, which was derived by Grabowska and Kaplan.
			They proposed a lattice formulation of four-dimensional chiral gauge theory, which is derived from their domain-wall formulation.
			In this formulation, an extra dimension is introduced and the gauge field along the extra dimension is evolved by the gradient flow.
			The chiral overlap operator satisfies the Ginsparg-Wilson relation and only depends on the gauge fields on the two boundaries.
			In this paper, we start from the arbitrary even-dimensional chiral overlap operator.
			We treat the gauge fields on the two boundaries independently, and derive the general expression to calculate the gauge anomaly with the chiral overlap operator in the continuum limit.
			As a result, we show that the gauge anomalies with the chiral overlap operator in two, four, and six dimensions in the continuum limit is equivalent to those known in the continuum theory up to total derivatives.
		\end{abstract}
		
	\end{titlepage}
	
	\setcounter{page}{1}
	
	\section{Introduction}
	It has been a long-standing problem to construct a gauge-invariant regularization for a chiral gauge theory.
	Grabowska and Kaplan proposed a formulation of the chiral gauge theory on the lattice~\cite{Grabowska:2015qpk}, which is developed based on the idea of the domain-wall fermion proposed by Kaplan~\cite{Kaplan:1992bt}.
	In the formulation of the domain-wall fermions, an extra dimension is introduced and the left-handed fermion is localized on one domain wall and the right-handed fermion is localized on the other domain wall.
	In this formulation, the left- and right-handed fermions are coupled with the same gauge field because the gauge field is constant along the extra dimension.
	 Thus this formulation is vector-like.
	On the other hand, in the Grabowska-Kaplan formulation the gauge field along the extra dimension is given by the gradient flow\cite{Narayanan:2006rf,Luscher:2009eq,Luscher:2010iy,Luscher:2011bx},
	\begin{equation}
		\partial_{s}\mathcal{A}_{\mu} = - \mathcal{D}_{\nu}\mathcal{F}_{\nu\mu},\quad \mathcal{A}_{\mu}(x,0) = A_{\mu}(x), \label{eq:flow}
	\end{equation} 
	where $\mathcal{A}_{\mu}(x,s)$ is the solution of the flow equation~\eqref{eq:flow}, and $\mathcal{D}_{\mu} = \partial_{\mu} + [\mathcal{A}_{\mu},\ \cdot \ ]$  and $\mathcal{F}_{\mu\nu}= \partial_{\mu}\mathcal{A}_{\nu} - \partial_{\nu}\mathcal{A}_{\mu} + [\mathcal{A}_{\mu},\mathcal{A}_{\nu}]$ are the covariant derivative and the field strength constructed from $\mathcal{A}_{\mu}(x,s)$, respectively.
	In other words, the gauge field is modified along the extra dimension by the gradient flow from the gauge field $A$ on one domain wall to $A_{\star}$ on the other domain wall.
	This means that the left- and right-handed fermions are coupled with the gauge field differently.
	Thus their formulation is expected to be a non-perturbative formulation of a chiral gauge theory.
	Grabowska and Kaplan also formulated a four-dimensional effective theory from the formulation explained above and obtained the chiral overlap operator~\cite{Grabowska:2016bis}, which obeys the Ginsparg-Wilson relation~\cite{Ginsparg:1981bj}.
	This operator only depends on the gauge fields $A$ and $A_{\star}$, and in the tree-level continuum limit, the left-handed fermion is only coupled with $A$ and the right-handed fermion is only coupled with $A_{\star}$.
	For the recent works related to Refs.~\cite{Grabowska:2015qpk,Grabowska:2016bis}, see Refs.~\cite{Fukaya:2016ofi,Okumura:2016dsr,Makino:2016auf,Makino:2017pbq,Hamada:2017tny}.
	
	The effective action of the $(2n+1)$-dimensional domain-wall fermion with the gauge field evolved by the gradient flow is composed of three parts: the effective action of the $2n$-dimensional left-handed fermion, the effective action of the $2n$-dimensional right-handed fermion, and the Chern-Simons term which is induced by the heavy modes in the bulk.
	Since the gradient flow assures the gauge invariance of the theory, the effective action of the $(2n+1)$-dimensional domain wall fermion itself is gauge-invariant.
	However, the effective actions of the left- and right-handed fermions are not gauge invariant because of gauge anomalies.
	In other words, the Chern-Simons term plays a role of cancelling out the gauge variation from the effective actions of the boundary modes.
	
	If the formulation is free from gauge anomalies, the Chern-Simons term vanishes.
	Moreover, as shown in the two-dimensional $\mathrm{U}(1)$ gauge theory in Refs.~\cite{Grabowska:2015qpk,Grabowska:2016bis}, the gauge field is expected to be evolved into a pure gauge so that the right-handed fermion on the other domain wall does not interact with the physical degrees of freedom of the gauge field. 
	Thus we expect that the $(2n+1)$-dimensional domain-wall fermion with the gauge field evolved by the gradient flow results in a $2n$-dimensional effective theory in which only the left-handed fermion couples to the physical degrees of freedom of the gauge field.
	
	In the lattice theory, we expect the same structure of the effective action in the continuum limit.
	The effective action constructed from the chiral overlap operator is composed of three parts: the functional of the gauge field $A$, the functional of the gauge field $A_{\star}$, and the cross terms of the gauge fields $A$ and $A_{\star}$.
	This effective action is gauge-invariant under the simultaneous gauge transformation of $A$ and $A_{\star}$.
	In the case of four-dimensional effective theories, the cross terms of the gauge fields $A$ and $A_{\star}$ were calculated in the continuum limit and it was confirmed that the parity-odd part of the gauge variation of the functional of the gauge field $A$ coincides with the gauge anomaly known in the continuum theory~\cite{Makino:2017pbq}.
	
	In order to confirm the correspondence of the structures of the effective actions between the formulation of the domain-wall fermion and the chiral overlap operator, we generalize this result; i.e., we calculate the gauge variation of the functional of the gauge field $A$ for an arbitrary even-dimensional effective action of the chiral overlap operator in the continuum limit, and explicitly check that the parity-odd part indeed coincides with the gauge anomaly in the continuum theory in the case of two, four, and six dimensions.
	
	\section{Notation and convention}    
	In Ref.~\cite{Grabowska:2016bis}, the chiral overlap operator is defined through the transfer matrix which depends on the flow time due to the $s$-dependence of the gauge field.
	They consider the simplification that the gauge field is constant in the half of the interval $[0, L]$ near the $s=0$ boundary and is $A_{\star}$ in the remaining region.
	$L$ is the length of the extra dimension and in the large $L$ limit the effective theory for the boundary modes is obtained.
	In this case, the chiral overlap operator in arbitrary even dimensions is expressed as follows:
	\begin{equation}
		a\hat{D}_{\chi} = 1+\gamma_{d+1}\left[1-(1-\epsilon_{\star})\frac{1}{1+\epsilon\epsilon_{\star}}(1-\epsilon)\right].
	\end{equation}
	Here $\epsilon$ is the sign function,
	\begin{equation}
		\epsilon = H_{\mathrm{W}}[A]\left(H_{\mathrm{W}}[A]^2\right)^{-1/2}, \label{eq:epsilon}
	\end{equation}
	of the Hermitian Wilson Dirac operator, 
	\begin{equation}
		H_{\mathrm{W}}[A] = \gamma_{d+1} \left\{\frac{1}{2}\left[\sum_{\mu}\gamma_{\mu}(\nabla^{*}_{\mu}[A] + \nabla_{\mu}[A]) -ar\sum_{\mu} \nabla^{*}_{\mu}[A]\nabla_{\mu}[A]\right] - M_0/a \right\}, \label{eq:HWDO}
	\end{equation}
	where $a$ is the lattice spacing, and $M_0$ and $r$ are free parameters.
	$\epsilon_{\star}$ is also defined by replacing the gauge field $A$ with $A_{\star}$ which is obtained from the original gauge field $A$ according to the gradient flow equation~\eqref{eq:flow}.
	Here we consider Euclidean arbitrary even dimensions $d=2n$.
	Gamma matrices satisfy the following equations:
	\begin{equation}
		\gamma_{\mu}^{\dagger} = \gamma_{\mu},\quad \{ \gamma_{\mu}, \gamma_{\nu} \} = 2\delta_{\mu \nu},\quad \gamma_{d+1} = i^n \gamma_1 \cdots \gamma_{d}. \label{def:gamma}
	\end{equation}
	Greek letters, $\mu,\nu,\ldots$, run from $1$ to $2n$. 
	Therefore,
	\begin{equation}
		\gamma_{d+1}^{\dagger} = \gamma_{d+1},\quad (\gamma_{d+1})^2 = 1,\quad  \tr \gamma_{d+1}\gamma_{\mu_1}\cdots \gamma_{\mu_d} = (-i)^n2^n\epsilon_{\mu_1 \cdots \mu_{d}}
	\end{equation}
	follow from Eq.~\eqref{def:gamma}. 
	$\nabla_{\mu}$ and $\nabla^{*}_{\mu}$ are the forward and backward lattice covariant derivatives, respectively, which are defined as 
	\begin{align}
		\nabla_{\mu}[A]f(x) &= \frac{1}{a}\left[ U(x,\mu)[A]f(x+a\hat{\mu})- f(x) \right], \\
		\nabla^{*}_{\mu}[A]f(x) &= \frac{1}{a}\left[ f(x)-U^{\dagger}[A](x-a\hat{\mu},\mu) f(x-a\hat{\mu}) \right]. 
	\end{align}
	The generators $T^a\, (a=1,\ldots, \mathop{\mathrm{dim}}\nolimits \mathcal{G})$ of the gauge group $\mathcal{G}$ satisfy the following equations:
	\begin{equation}
		(T^a)^{\dagger}=-T^a,\quad [T^a,T^b] = f^{abc}T^c,\quad \tr T^aT^b = -1/2\delta^{ab}. 
	\end{equation}
	Here the covariant derivative is defined as $D_{\mu}=\partial_{\mu} + A_{\mu}$, where $A_{\mu} = A^a_{\mu}T^a$ and $A^a_{\mu}$ is real.
	Thus by defining the link variable as
	\begin{equation}
		U(x,\mu)[A] = \mathcal{P}\exp\left[ a\int_0^1 \diff t\, A_{\mu}(x+ta\hat{\mu}) \right],
	\end{equation}
	where $\mathcal{P}$ denotes the path-ordered product and $\hat{\mu}$ is the unit vector in the direction $\mu$, 
	we obtain 
	\begin{align}
		\nabla_{\mu}[A]f(x) &= (D_{\mu} + \mathcal{O}(a))f(x), \\
		\nabla^{*}_{\mu}[A]f(x) &= (D_{\mu} + \mathcal{O}(a))f(x),
	\end{align}
	in the continuum limit $a \rightarrow 0$.
	In Ref.~\cite{Makino:2017pbq}, the fermion one-loop effective action defined by
	\begin{equation}
		\Gamma_{\mathrm{lat.}}[A,A_{\star}]\equiv - \ln \int \prod_{x} [\diff \psi(x)\diff \bar{\psi}(x)]\, \exp \left[ -a^d\sum_{x}\bar{\psi}(x)\hat{D}_{\chi}\psi(x) \right]
	\end{equation}
	is studied and the following expression is obtained:
	\begin{equation}
		\delta\delta_{\star} \Gamma_{\mathrm{lat.}}[A,A_{\star}] = -\frac{1}{2}\Tr ( 1 - \epsilon_{\star} )\frac{1}{\epsilon + \epsilon_{\star}}\delta \epsilon \frac{1}{\epsilon + \epsilon_{\star}}\delta_{\star}\epsilon_{\star}, \label{eq:DVoEA}
	\end{equation}
	where $\Tr \equiv \sum_x \tr$ and $\tr$ denotes the trace over the spinor and gauge indices.
	$\delta$ and $\delta_{\star}$ are the infinitesimal variations which only act on $A$ and $A_{\star}$, respectively:
	\begin{align}
		\delta A &\not= 0,\quad \delta A_{\star} = 0, \\
		\delta_{\star} A_{\star} &\not= 0,\quad \delta_{\star} A = 0.
	\end{align}
	Here, we treat the gauge fields $A$ and $A_{\star}$ independently, and the infinitesimal variations $\delta$ and $\delta_{\star}$ are independently.\footnote{We also assume that $A$ and $A_{\star}$ have the same winding number so that $\epsilon + \epsilon_{\star}$ does not have zero eigenvalues (see Appendix in Ref.~\cite{Grabowska:2016bis}). }
	Eq.~\eqref{eq:DVoEA} is decomposed into the parity-odd and parity-even parts. The former is written as
	\begin{align}
		(\text{parity-odd part}) &= \frac{1}{2}\Tr\epsilon_{\star}\frac{1}{\epsilon+\epsilon_{\star}}\delta\epsilon\frac{1}{\epsilon+\epsilon_{\star}}\delta_{\star}\epsilon_{\star} \\
		&=-\frac{1}{2}\delta\left( \Tr\epsilon_{\star}\frac{1}{\epsilon+\epsilon_{\star}}\delta_{\star}\epsilon_{\star} \right),
	\end{align}
	and the latter is written as
	\begin{align}
		(\text{parity-even part}) &= -\frac{1}{2}\Tr\frac{1}{\epsilon+\epsilon_{\star}}\delta\epsilon\frac{1}{\epsilon+\epsilon_{\star}}\delta_{\star}\epsilon_{\star} \\
		&=\frac{1}{2}\delta\delta_{\star}\Tr\ln(\epsilon +\epsilon_{\star}).
	\end{align}
	By expressing the infinitesimal gauge transformation as
	\begin{alignat}{2}
		\delta^{\omega} A_{\mu}(x) &= \partial_{\mu}\omega(x) + [ A_{\mu}(x) , \omega(x) ], &\qquad  \delta^{\omega} {A_{\star}}_{\mu}(x) &= 0, \label{eq:gaugetrans_A} \\
		\delta^{\omega}_{\star} {A_{\star}}_{\mu}(x) &= \partial_{\mu}\omega(x) + [ {A_{\star}}_{\mu}(x) , \omega(x) ], & \delta^{\omega}_{\star} A_{\mu}(x) &= 0, \label{eq:gaugetrans_Astar}
	\end{alignat}
	and using the equations,
	\begin{align}
		(\delta^{\omega}+\delta_{\star}^{\omega})\Gamma_{\mathrm{lat.}}[A,A_{\star}] &= 0, \\
		(\delta^{\omega}+\delta_{\star}^{\omega})\Tr\ln(\epsilon +\epsilon_{\star}) &= 0,
	\end{align}
	we obtain the following equation which is related to the gauge anomaly:
	\begin{equation}
		\delta^{\omega}\Gamma_{\mathrm{lat.}}[A,0] = \frac{1}{2} \Tr\epsilon_{\star}\frac{1}{\epsilon+\epsilon_{\star}}\delta_{\star}^{\omega}\epsilon_{\star}[A,0] + \frac{1}{2}\delta^{\omega} \Tr\ln(\epsilon +\epsilon_{\star})[A,0]. \label{eq:Gamma_lat}
	\end{equation}
	The parity-even part can be removed by local counterterms.
	As discussed in Ref.~\cite{Makino:2017pbq} for $d=4$, the parity-even part contains a mass term of the gauge field even if the anomaly free condition is satisfied, and this part should be subtracted by local counterterms.
		
	\section{Calculation of the gauge anomaly}
	In this section, we evaluate the parity-odd part following Ref.~\cite{Fujiwara:2002xh}, in which the axial anomaly  $-1/(2a^d)\tr \epsilon (x,x)$ defined by the overlap operator is calculated in the continuum limit for arbitrary even dimensions.
	By expanding the parity-odd part of Eq.~\eqref{eq:Gamma_lat} in powers of $\Delta \equiv \epsilon - \epsilon_{\star}= \mathcal{O}(a)$, we obtain the following equations:
	\begin{align}
		\frac{1}{2} \Tr\epsilon_{\star}\frac{1}{\epsilon+\epsilon_{\star}}\delta_{\star}^{\omega}\epsilon_{\star}[A,0]
		&= \left.a^d\sum_{x}\frac{1}{2a^d}\tr \epsilon_{\star}(\epsilon+\epsilon_{\star})\frac{1}{(\epsilon+\epsilon_{\star})^2}\delta_{\star}^{\omega}\epsilon_{\star}\right|_{A_{\star}=0}(x,x) \nonumber \\
		&= \left.a^d\sum_{x}\frac{1}{2a^d}\tr \epsilon_{\star}(2\epsilon_{\star}+\Delta)\frac{1}{4-\Delta^2}\delta_{\star}^{\omega}\epsilon_{\star}\right|_{A_{\star}=0}(x,x) \nonumber \\
		&= \left.a^d\sum_{x}\frac{1}{4a^d}\tr \left[\sum_{\ell=0}^{\infty}(\Delta/2)^{2\ell}+\epsilon_{\star}\sum_{\ell=0}^{\infty}(\Delta/2)^{2\ell+1}\right]\delta_{\star}^{\omega}\epsilon_{\star}\right|_{A_{\star}=0}(x,x). 
	\end{align}
	Since $a^d \sum_x \rightarrow \int \diff^d x$, we only need to calculate $\mathcal{O}(a^m)$ terms with $m \leq d$ in the trace of 
	\begin{equation}
		\mathcal{A}_{\mathrm{gauge}}(x) \equiv \left.\frac{1}{4a^d}\tr \left[\sum_{\ell=0}^{\infty}(\Delta/2)^{2\ell}+\epsilon_{\star}\sum_{\ell=0}^{\infty}(\Delta/2)^{2\ell+1}\right]\delta_{\star}^{\omega}\epsilon_{\star}\right|_{A_{\star}=0}(x,x). \label{eq:gauge-anomaly}
	\end{equation}
	From Eqs.~\eqref{eq:epsilon} and \eqref{eq:HWDO}, it is clear that $\epsilon$ and $\epsilon_{\star}$ contain one $\gamma_{d+1}$. Thus $\gamma_{d+1}$s appear odd times in all of the terms of Eq.~\eqref{eq:gauge-anomaly} and these terms are reduced to the form that contains the factor $\tr \gamma_{d+1}\gamma_{\mu_1}\cdots \gamma_{\mu_m}$, which is zero if $m < d$.
	Therefore, we only need to take account of the terms in which $\gamma_{\mu}$s appear at least $d$ times.
	Note that a diagonal element of the kernel of an operator $\mathcal{O}$ on the lattice is calculated from
	\begin{equation}
		\mathcal{O}(x,x) = \int_{\mathcal{B}^d} \frac{\diff^d k}{(2\pi)^d}\, e^{-ikx/a}(\mathcal{O}e^{ikx/a}),
	\end{equation}
	where
	\begin{equation}
		\mathcal{B}^d \equiv \left\{\, (k_1, \ldots, k_d)\in \mathbb{R}^d \, \big|\,  -\pi \leq k_{\mu} \leq \pi,\ \forall \mu \in \{1,\ldots,d\} \,\right\}.
	\end{equation}
	From Eqs.~\eqref{eq:epsilon} and \eqref{eq:HWDO}, we obtain 
	\begin{equation}
		\epsilon e^{ikx/a} f(x) =e^{ikx/a} \gamma_{d+1}\left[(V + aS^{-1}P_1)\sum_{\ell=0}^{\infty} a^{2\ell}\alpha_{2\ell}S^{-2\ell}P_2^{\ell} + \cdots \right]f(x), \label{eq:epsilon_obtained}
	\end{equation}
	where the ellipsis denotes the terms which do not contribute to Eq.~\eqref{eq:gauge-anomaly} in the continuum limit for the reason explained later.
	The expressions which appear in Eq.~\eqref{eq:epsilon_obtained} are defined as follows:
	\begin{align}
		S &= \left( \sum_{\nu}s_{\nu}^2 + M^2 \right)^{1/2}, \\
		V &= \left(\sum_{\mu}\gamma_{\mu}is_{\mu} - M \right)S^{-1}, \\
		P_1 &= \sum_{\mu}\gamma_{\mu} c_{\mu}D_{\mu} - r\sum_{\mu} is_{\mu}D_{\mu}, \label{eq:P_1}\\
		P_2 &= \frac{1}{2}\sum_{\nu,\rho}\gamma_{\nu}\gamma_{\rho}c_{\nu}c_{\rho}F_{\nu\rho} - r\sum_{\nu,\rho}\gamma_{\nu}c_{\nu} is_{\rho}F_{\nu\rho}, \label{eq:P_2}
	\end{align}
	where
	\begin{equation}
		s_{\mu} = \sin k_{\mu},\quad c_{\mu} = \cos k_{\mu},
	\end{equation}
	\begin{equation}
		M = M_0 + r\sum_{\rho}(c_{\rho}-1),
	\end{equation}
	and $F_{\mu\nu} = [D_{\mu},D_{\nu}]=\partial_{\mu}A_{\nu}-\partial_{\nu}A_{\mu} + [A_{\mu},A_{\nu}]$ denotes the field strength of the gauge field $A_{\mu}(x)$.
	$\alpha_{2\ell}$ is defined as the coefficient of $z^{2\ell}$ in the power series of $(1-z^2)^{-1/2}$; that is, $(1-z^2)^{-1/2} = \sum_{\ell} \alpha_{2\ell}z^{2\ell}$, and the explicit form of $\alpha_{2\ell}$ is given as follows:
	\begin{equation}
		\alpha_{2\ell} = \frac{1}{\ell!} \frac{\Gamma(\ell+1/2)}{\Gamma(1/2)}.
	\end{equation}
	Since we have $P_2=0$ if $A_{\mu}=0$, we obtain the following expressions:
	\begin{align}
		\left.\epsilon_{\star}\right|_{A_{\star}=0}e^{ikx/a} f(x) &= e^{ikx/a}\gamma_{d+1}(V + aS^{-1}P_1|_{A=0} + \cdots)f(x), \label{eq:epsilon-star} \\
		\left.\delta_{\star}^{\omega}\epsilon_{\star}\right|_{A_{\star}=0}e^{ikx/a} f(x) &= e^{ikx/a}\gamma_{d+1} (aS^{-1}\delta^{\omega} P_1|_{A=0} + \cdots)f(x), \label{eq:delta-epsilon-star}
	\end{align}
	and
	\begin{align}
		&\left.\Delta \right|_{A_{\star}=0}e^{ikx/a}f(x) \nonumber\\
		&=e^{ikx/a} \gamma_{d+1}\left[aS^{-1}( P_1-P_1|_{A=0}) +(V + aS^{-1}P_1) \sum_{\ell=1}^{\infty} a^{2\ell}\alpha_{2\ell}S^{-2\ell}P_2^{\ell}+ \cdots\right] f(x). \label{eq:Delta}
	\end{align}
	The number of $\gamma_{\mu}$s for each of the $a^{m}$-terms in Eqs.~\eqref{eq:epsilon-star}, \eqref{eq:delta-epsilon-star}, and \eqref{eq:Delta} is less than or equal to $m$ except for the terms which contain $V$s. 
	On the other hand, the maximum number of $\gamma_{\mu}$s for the $a^{m}$-terms which contain $V$s is $m+1$. 
	Therefore, in the $a^{p}$-terms in Eq.~\eqref{eq:gauge-anomaly} where $V$s appear $q$ times, $\gamma_{\mu}$s appear at most $p+q$ times. 
	However, the number of $\gamma_{\mu}$s can be reduced as we explain below.
	
	From the equation,
	\begin{equation}
		\gamma_{\nu}(\gamma_{\mu_1}\cdots \gamma_{\mu_m}) = \sum_j (-1)^{j-1}2\delta_{\nu\mu_j}\gamma_{\mu_1}\cdots \hat{\gamma}_{\mu_j}\cdots \gamma_{\mu_m} + (-1)^m(\gamma_{\mu_1}\cdots \gamma_{\mu_m})\gamma_{\nu},
	\end{equation}
	we obtain the following equations:
	\begin{align}
		\gamma_{d+1}V(\gamma_{\mu_1} \cdots \gamma_{\mu_{2m'}})\gamma_{d+1}V = \gamma_{\mu_{1}} \cdots \gamma_{\mu_{2m'}} +\sum_{j}(-1)^{j}2 is_{\mu_j}(\gamma_{\mu_1} \cdots \hat{\gamma}_{\mu_j} \cdots \gamma_{\mu_{2m'}})S^{-1} V, \label{eq:V-gamma} \\
		V(\gamma_{\mu_1} \cdots \gamma_{\mu_{2m'-1}})V = \gamma_{\mu_{1}} \cdots \gamma_{\mu_{2m'-1}} +\sum_{j}(-1)^{j-1}2 is_{\mu_j}(\gamma_{\mu_1} \cdots \hat{\gamma}_{\mu_j} \cdots \gamma_{\mu_{2m'-1}}) S^{-1} V. \label{eq:V-gamma_odd} 
	\end{align}
	Here $\hat{\gamma}_{\mu_j}$ means that $\gamma_{\mu_j}$ is omitted.
	From Eqs.~\eqref{eq:V-gamma} and \eqref{eq:V-gamma_odd}, the $a^{m}$-terms in Eq.~\eqref{eq:gauge-anomaly} are reduced to the form in which the number of $\gamma_{\mu}$s is less than or equal to $m+1$. 
	In addition, since the maximum numbers of $\gamma_{\mu}$s in $P_1P_2^{m}$ and $VP_2^m$ are odd, and the terms in Eq.~\eqref{eq:gauge-anomaly} are the products of odd number of them, the maximum number of $\gamma_{\mu}$s in the $a^{d-1}$-terms in Eq.~\eqref{eq:gauge-anomaly} is odd, that is, not $d$ but $d-1$.
	Therefore, we only need to consider the $a^{d}$-terms in Eq.~\eqref{eq:gauge-anomaly} in the continuum limit.
	
	Let $N_V$ be the number of $V$s in each of the $a^{d}$-terms and $N_{\gamma}$ is the number of $\gamma_{\mu}$s without $\gamma_{\mu}$s in $V$s in each of the $a^{d}$-terms. Then each of the $a^{d}$-terms in Eq.~\eqref{eq:gauge-anomaly} is classified into six cases.
	
	\vspace{.5\baselineskip}
	\noindent(i) $N_V$ is odd and $N_{\gamma}=d$. 
	\begin{quote}
	The terms classified into the case (i) are the products of $V$s and the first terms of the right-hand sides of Eqs.~\eqref{eq:P_1} and \eqref{eq:P_2}.
	From Eqs.~\eqref{eq:V-gamma} and \eqref{eq:V-gamma_odd}, the number of $\gamma_{\mu}$s is reduced to $d+1$ at most.
	\begin{equation}
		\tr (\text{odd $\gamma$s})V \cdots \underbrace{V(\text{odd $\gamma$s})V}_{\text{Eq.~\eqref{eq:V-gamma} or \eqref{eq:V-gamma_odd}}}(\text{odd $\gamma$s})\underbrace{V(\text{odd $\gamma$s})V}_{\text{Eq.~\eqref{eq:V-gamma} or \eqref{eq:V-gamma_odd}}} \cdots V(\text{even $\gamma$s})
	\end{equation}
	Since $\tr \gamma_{d+1}\gamma_{\mu_1}\cdots \gamma_{\mu_{d+1}} = 0$, the terms which contain $d$ $\gamma_{\mu}$s only contribute to Eq.~\eqref{eq:gauge-anomaly}.
	In the terms which contain the second term of Eq.~\eqref{eq:V-gamma} or \eqref{eq:V-gamma_odd}, $V$s appear at least twice and the total number of $\gamma_{\mu}$s and $\gamma_{d+1}$s between them is odd.
	Using Eq.~\eqref{eq:V-gamma} or \eqref{eq:V-gamma_odd}, one can reduce the number of $\gamma_{\mu}$s in these terms by two and only the first terms of Eq.~\eqref{eq:V-gamma} remain.
	Thus the terms classified into the case (i) are equivalent to the terms from which all $V$s are removed  and which is multiplied by $-M/S$ because one $V$ remains.
	\end{quote}
	
	(ii) $N_V$ is odd and $N_{\gamma}=d-1$. 
	\begin{quote}
	The terms classified into the case (ii) are the products of $V$s and the first terms of the right-hand sides of Eqs.~\eqref{eq:P_1} and \eqref{eq:P_2} except for one factor which is replaced with the second term.
	From Eqs.~\eqref{eq:V-gamma} and \eqref{eq:V-gamma_odd}, the number of $\gamma_{\mu}$s is reduced to $d$ at most.
	\begin{equation}
	 \tr (\text{odd $\gamma$s})V \cdots \underbrace{V(\text{odd $\gamma$s})V}_{\text{Eq.~\eqref{eq:V-gamma} or \eqref{eq:V-gamma_odd}}}(\text{odd $\gamma$s})V(\text{even $\gamma$s})\underbrace{V(\text{odd $\gamma$s})V}_{\text{Eq.~\eqref{eq:V-gamma} or \eqref{eq:V-gamma_odd}}} \cdots V(\text{even $\gamma$s})
	\end{equation}
	The second terms of Eqs.~\eqref{eq:V-gamma} and \eqref{eq:V-gamma_odd} do not contribute to Eq.~\eqref{eq:gauge-anomaly} as explained in the case (i).
	Moreover, only the part $\sum_{\nu}\gamma_{\nu}is_{\nu}/S$ in the remaining $V$ contribute to Eq.~\eqref{eq:gauge-anomaly}.
	The total number of $\gamma_{\mu}$s and $\gamma_{d+1}$s between the factor $\sum_{\nu}\gamma_{\nu}is_{\nu}/S$ and the factor with respect to the second term of the right-hand sides of Eqs.~\eqref{eq:P_1} and \eqref{eq:P_2} is odd for Eq.~\eqref{eq:P_1} and even for Eq.~\eqref{eq:P_2}.
 	Therefore, the terms classified into the case (ii) are equivalent to the terms from which all $V$s are removed and in which the factor $\mp \sum_{\nu}\gamma_{\nu}is_{\nu}/S$ is inserted before the factor with respect to the second term of the right-hand sides of Eqs.~\eqref{eq:P_1} and \eqref{eq:P_2} with the negative sign for Eq.~\eqref{eq:P_1} and the positive sign for Eq.~\eqref{eq:P_2}.

	\end{quote}
	(iii-a) $N_V$ is odd and $N_{\gamma}$ is odd with $N_{\gamma} \leq d-3$. \newline
	(iii-b) $N_V$ is even and $N_{\gamma}$ is even with $N_{\gamma} \leq d-2$.
	\begin{quote}
	From Eqs.~\eqref{eq:V-gamma} and \eqref{eq:V-gamma_odd}, the number of $\gamma_{\mu}$s is reduced to $d$ at most and at least two $V$s remain. Thus in the terms in which $\gamma_{\mu}$s appear $d$ times, the $\left(\sum_{\nu}\gamma_{\nu}s_{\nu}\right)$s from the remaining $V$s appear twice at least. Since $\tr \gamma_{d+1}\gamma_{\mu_1}\cdots \gamma_{\mu_d} = (-i)^n2^n\epsilon_{\mu_1 \cdots \mu_{d}}$ and $\epsilon_{\mu_1 \cdots \mu_{d}}$ is antisymmetric with respect to the subscripts, the terms classified into the cases (iii-a) and (iii-b) do not contribute to Eq.~\eqref{eq:gauge-anomaly}.
	\end{quote}
	(iv-a) $N_V$ is odd and $N_{\gamma}$ is even with $N_{\gamma} \leq d-2$. \newline
	(iv-b) $N_V$ is even and $N_{\gamma}$ is odd with $N_{\gamma} \leq d-1$.  
	\begin{quote}
	From Eqs.~\eqref{eq:V-gamma} and \eqref{eq:V-gamma_odd}, the number of $\gamma_{\mu}$s is reduced to $d-1$ at most. Thus the terms classified into the cases (iv-a) and (iv-b) do not contribute to Eq.~\eqref{eq:gauge-anomaly}.
	\end{quote}
	
	Therefore, the terms classified into the cases (i) and (ii) only contribute to Eq.~\eqref{eq:gauge-anomaly} in the continuum limit.
	Note that the terms which contain the factors of ellipses in Eqs.~\eqref{eq:epsilon-star}, \eqref{eq:delta-epsilon-star}, and \eqref{eq:Delta} do not contribute to Eq.~\eqref{eq:gauge-anomaly} in the continuum limit, because they correspond to the cases (iii-a), (iii-b), (iv-a), and (iv-b).
	
	Now, we can write down the terms which contribute to Eq.~\eqref{eq:gauge-anomaly} in the continuum limit.
	Here we define the following functions:
	\begin{align}
		\tilde{D}(\lambda) &=  \sum_{\mu}\gamma_{\mu} c_{\mu}D_{\mu}-\lambda \left(\sum_{\mu}\gamma_{\mu}is_{\mu} \right)\left(- r\sum_{\mu} is_{\mu}D_{\mu}\right), \\
		\tilde{F}(\lambda) &= \frac{1}{2}\sum_{\nu,\rho}\gamma_{\nu}\gamma_{\rho}c_{\nu}c_{\rho}F_{\nu\rho}+\lambda \left(\sum_{\mu}\gamma_{\mu}is_{\mu}\right)\left(-r\sum_{\nu,\rho}\gamma_{\nu}c_{\nu} is_{\rho}F_{\nu\rho}\right),
	\end{align}
	and
	\begin{equation}
		\left\{
		\begin{aligned}
		\tilde{P}_1(\lambda) &= \tilde{D}(\lambda) - \tilde{D}^0(\lambda), \\
		\tilde{P}_{2\ell}(\lambda) &= \alpha_{2\ell}\tilde{F}(\lambda)^{\ell}, \\
		\tilde{P}_{2\ell+1}(\lambda) &=  \alpha_{2\ell}\tilde{D}(\lambda)\tilde{F}(\lambda)^{\ell},
		\end{aligned}
		\right.
	\end{equation}
	where $\tilde{D}^0(\lambda)= \tilde{D}(\lambda)|_{A=0}$.
	Then, we obtain 
	\begin{align}
		&e^{-ikx/a}\left(\frac{1}{a^d}\left. \tr \Delta^{2\ell}\delta_{\star}^{\omega}\epsilon_{\star}\right|_{A_{\star}=0}\right)e^{ikx/a} \nonumber \\
		&=S^{-2n-1}\tr \sum_{\sum\limits_{m=1}^{2\ell}i_m=d-1} \Biggl( (-M)( \gamma_{d+1}\tilde{P}_{i_1})\cdots (\gamma_{d+1}\tilde{P}_{i_{2\ell}})(\gamma_{d+1}\delta^{\omega}\tilde{P}_1) \nonumber \\
		&\hspace{4cm}+ \frac{\diff}{\diff \lambda}(\gamma_{d+1}\tilde{P}_{i_1})\cdots (\gamma_{d+1}\tilde{P}_{i_{2\ell}})(\gamma_{d+1}\delta^{\omega}\tilde{P}_1) \Biggr)\Biggr|_{\lambda = 0} + \mathcal{O}(a) \nonumber \\
		&=-S^{-2n-1}\tr \sum_{\sum\limits_{m=1}^{2\ell}i_m=d-1}(-1)^{\sum\limits_{m'=1}^{\ell}i_{2m'}}\Biggl( M\gamma_{d+1}\tilde{P}_{i_1}\cdots \tilde{P}_{i_{2\ell}}\delta^{\omega}\tilde{P}_1\nonumber \\
		&\hspace{6.1cm}- \gamma_{d+1}\frac{\diff}{\diff \lambda}\tilde{P}_{i_1}\cdots \tilde{P}_{i_{2\ell}} \delta^{\omega}\tilde{P}_1 \Biggr)\Biggr|_{\lambda = 0} + \mathcal{O}(a), \label{eq:Delta^2l}
	\end{align}
	and
	\begin{align}
		&e^{-ikx/a}\left(\frac{1}{a^d}\left. \tr \epsilon_{\star}\Delta^{2\ell+1}\delta_{\star}^{\omega}\epsilon_{\star}\right|_{A_{\star}=0}\right)e^{ikx/a} \nonumber \\
		&=S^{-2n-1} \tr \sum_{\sum\limits_{m=1}^{2\ell+1}i_m=d-1} \Biggl((-M)\gamma_{d+1}( \gamma_{d+1}\tilde{P}_{i_1})\cdots (\gamma_{d+1}\tilde{P}_{i_{2\ell+1}}) (\gamma_{d+1}\delta^{\omega}\tilde{P}_1) \nonumber\\
		&\hspace{4cm}+ \frac{\diff}{\diff \lambda}\gamma_{d+1}(\gamma_{d+1}\tilde{P}_{i_1})\cdots (\gamma_{d+1}\tilde{P}_{i_{2\ell+1}}) (\gamma_{d+1}\delta^{\omega}\tilde{P}_1)\Biggr)\Biggr|_{\lambda=0}\nonumber\\
		&\quad+S^{-2n-1} \tr \sum_{\sum\limits_{m=1}^{2\ell+1}i_m=d-2}\Biggl( (-M)(\gamma_{d+1}\tilde{D}^0) (\gamma_{d+1}\tilde{P}_{i_1}) \cdots (\gamma_{d+1}\tilde{P}_{i_{2\ell+1}}) (\gamma_{d+1}\delta^{\omega}\tilde{P}_1) \nonumber \\
		&\hspace{3cm}+ \frac{\diff}{\diff \lambda} (\gamma_{d+1}\tilde{D}^0) (\gamma_{d+1}\tilde{P}_{i_1}) \cdots (\gamma_{d+1}\tilde{P}_{i_{2\ell+1}}) (\gamma_{d+1}\delta^{\omega}\tilde{P}_1) \Biggr)\Biggr|_{\lambda=0}+ \mathcal{O}(a) \nonumber \\
		&=-S^{-2n-1} \tr \sum_{\sum\limits_{m=1}^{2\ell+1}i_m=d-2} (-1)^{\sum\limits_{m'=0}^{\ell}i_{2m'+1}}\Biggl(M\gamma_{d+1}\tilde{P}_{i_1}\cdots \tilde{P}_{i_{2\ell+1}}\delta^{\omega}\tilde{P}_1 \nonumber\\
		&\hspace{6.4cm}- \gamma_{d+1}\frac{\diff}{\diff \lambda}\tilde{P}_{i_1}\cdots \tilde{P}_{i_{2\ell+1}}\delta^{\omega}\tilde{P}_1\Biggr)\Biggr|_{\lambda=0}\nonumber\\
		&\quad -S^{-2n-1} \tr \sum_{\sum\limits_{m=1}^{2\ell+1}i_m=d-2}(-1)^{\sum\limits_{m'=0}^{\ell}i_{2m'+1}}\Biggl( M\gamma_{d+1}\tilde{D}^0\tilde{P}_{i_1}\cdots \tilde{P}_{i_{2\ell+1}}\delta^{\omega}\tilde{P}_1 \nonumber \\
		&\hspace{6.4cm}- \gamma_{d+1}\frac{\diff}{\diff \lambda}\tilde{D}^0\tilde{P}_{i_1}\cdots \tilde{P}_{i_{2\ell+1}}\delta^{\omega}\tilde{P}_1\Biggr)\Biggr|_{\lambda=0}+ \mathcal{O}(a). \label{eq:Delta^2l+1}
	\end{align}
	Eqs.~\eqref{eq:Delta^2l} and \eqref{eq:Delta^2l+1} can be simplified further by evaluating the trace over the spinor index and the integral with respect to the variable $k$.
	In general, by defining
	\begin{align}
		X_1 &\equiv M\sum_{\mu_1,\ldots, \mu_{2n}} \gamma_{\mu_1}\cdots\gamma_{\mu_{2n}}c_{\mu_1}\cdots c_{\mu_{2n}}X_{\mu_1 \cdots \mu_{2n}}, \\
		X_2 &\equiv r\sum_{i} (-1)^{i-1}\sum_{\sigma}\gamma_{\sigma}s_{\sigma}\sum_{\mu_1,\ldots, \mu_{2n}}\gamma_{\mu_1}\cdots\hat{\gamma}_{\mu_i}\cdots \gamma_{\mu_{2n}} s_{\mu_i}c_{\mu_1}\cdots \hat{c}_{\mu_{i}}\cdots c_{\mu_{2n}}X_{\mu_1 \cdots \mu_{2n}},
	\end{align}
	where $X_{\mu_1 \cdots \mu_{2n}}$ is independent of $k_{\mu}$ and valued in the Lie algebra of the gauge group $\mathcal{G}$ with $2n$ subscripts running from $1$ to $2n$,
	we have the following equations:
	\begin{align}
		&\int_{\mathcal{B}^d} \frac{\diff^d k}{(2\pi)^d}\, S^{-2n-1} \left[ \tr \left( \gamma_{d+1}X_1 \right) +\tr \left( \gamma_{d+1}X_2 \right)\right] \nonumber \\
		&=\int_{\mathcal{B}^d} \frac{\diff^d k}{(2\pi)^d}\, S^{-2n-1} (-i)^n2^n\sum_{\mu_1,\ldots, \mu_{2n}}\epsilon_{\mu_1 \cdots \mu_{2n}}c_{\mu_1}\cdots c_{\mu_{2n}}\left( M +r\sum_{i}s_{\mu_i}^2/c_{\mu_i}\right) X_{\mu_1 \cdots \mu_{2n}} \nonumber \\
		&=\int_{\mathcal{B}^d} \frac{\diff^d k}{(2\pi)^d}\, S^{-2n-1} (-i)^n2^n\left(\prod_{\mu}c_{\mu}\right)\left( M +r\sum_{\rho}s_{\rho}^2/c_{\rho}\right)\sum_{\mu_1,\ldots, \mu_{2n}}\epsilon_{\mu_1 \cdots \mu_{2n}} X_{\mu_1 \cdots \mu_{2n}} \nonumber \\
		&=\frac{2(-i)^n}{(2\pi)^n}\frac{\Gamma(1/2)}{\Gamma(n+1/2)}I(M_0,r)\sum_{\mu_1,\ldots ,\mu_{2n}}\epsilon_{\mu_1 \cdots \mu_{2n}}X_{\mu_1\cdots \mu_{2n}} \nonumber \\
		&= \frac{(-i)^n}{(2\pi)^nn!} \frac{2}{\alpha_{2n}}\sum_{\mu_1,\ldots ,\mu_{2n}}\epsilon_{\mu_1 \cdots \mu_{2n}}X_{\mu_1\cdots \mu_{2n}},  \label{eq:X}
	\end{align}
	where $\tr$ denotes the trace over the spinor only, and
	\begin{align}
		I(M_0,r) &\equiv \frac{1}{2\pi^n}\frac{\Gamma(n+1/2)}{\Gamma(1/2)} \int_{\mathcal{B}^d}\diff^d k\, \left(\prod_{\mu}c_{\mu}\right) S^{-n-1/2} \left( M+r\sum_{\rho}s_{\rho}^2/c_{\rho} \right) \nonumber \\
		&= \sum_{n_{\pi}=0}^{\lfloor M_0/(2r)\rfloor} \frac{d!}{n_{\pi}!(d-n_{\pi})!} (-1)^{n_{\pi}}. \label{eq:derivation_of_I}
	\end{align}
	$\lfloor x \rfloor$ denotes the maximum integer which is less than or equal to $x$.
	For the derivation of Eq.~\eqref{eq:derivation_of_I}, see Ref.~\cite{Fujiwara:2002xh}.
	In the last equality of Eq.~\eqref{eq:X}, we used $I(M_0,r)=1$ by assuming $0<M_0/r<2$.
	Therefore, using Eqs.~\eqref{eq:Delta^2l}, \eqref{eq:Delta^2l+1} and \eqref{eq:X}, we finally obtain the following expression:
	\begin{equation}
		\lim_{a \to 0} a^d\sum_x \mathcal{A}_{\mathrm{gauge}}(x) = \frac{1}{n!}\left( \frac{-i}{2\pi}\right)^n \int \frac{-1}{2\alpha_{2n}}\tr G,
	\end{equation}
	where
	\begin{align}
		G &=\sum_{\ell=0}^{\infty}\left[\sum_{\sum\nolimits_{m=1}^{2\ell}i_m=d-1} (-1)^{\sum_{m'}i_{2m'}} (1/2)^{2\ell}g_{i_1}\cdots g_{i_{2\ell}} \right.\nonumber\\
		&\hspace{0.9cm}+ \sum_{\sum\nolimits_{m=1}^{2\ell+1}i_m=d-1} (-1)^{\sum_{m'}i_{2m'+1}}(1/2)^{2\ell+1}g_{i_1}\cdots g_{i_{2\ell+1}} \nonumber\\
		& \hspace{0.9cm}\left.+ \sum_{\sum\nolimits_{m=1}^{2\ell+1}i_m=d-2}  (-1)^{\sum_{m'}i_{2m'+1}}(1/2)^{2\ell+1} \diff g_{i_1}\cdots g_{i_{2\ell+1}} \right]\diff \omega, \label{eq:G}
	\end{align}
	and
	\begin{equation}
		\left\{
		\begin{aligned}
		g_1 &= A, \\
		g_{2m} &= \alpha_{2m}F^m, \\
		g_{2m+1} &= \alpha_{2m}DF^m.
		\end{aligned}
		\right.
	\end{equation}
	Here $\omega$ is defined by Eqs.~\eqref{eq:gaugetrans_A} and \eqref{eq:gaugetrans_Astar}, and we used the differential form. That is, $A= A_{\mu}\diff x^{\mu}$, $D=\diff +A$, and $F = D^2 = 1/2F_{\mu\nu}\diff x^{\mu} \wedge \diff x^{\nu}$.
	We can obtain the explicit form in the specific dimensions as follows:
	
	\begin{enumerate}
	\item $d=2$:
	
	From Eq.~\eqref{eq:G}, we have
	\begin{equation}
		G = -1/2A\diff \omega.
	\end{equation}
	Thus we obtain
	\begin{align}
		\lim_{a \to 0} a^2\sum_x \mathcal{A}_{\mathrm{gauge}}(x) =-\frac{i}{4\pi}\int \tr A\diff \omega.
	\end{align}
	
	\item $d=4$:
	
	From Eq.~\eqref{eq:G}, we have
	\begin{align}
		G &= (- 1/2 \alpha_2 DF+1/4 \alpha_2 A\cdot F -1/4 \alpha_2 F\cdot A +1/2 \alpha_2\diff F + 1/8 A\cdot A\cdot A)\diff \omega \nonumber \\
		&= -1/8(A(\diff A) + (\diff A) A + A^3)\diff \omega.
	\end{align}
	Thus we obtain
	\begin{align}
		\lim_{a \to 0} a^4\sum_x \mathcal{A}_{\mathrm{gauge}}(x) =-\frac{1}{48\pi^2}\int (A(\diff A) + (\diff A) A + A^3)\diff \omega.
	\end{align}
	Tis result is derived in Ref.~\cite{Makino:2017pbq}.
	
	\item $d=6$:
	
	From Eq.~\eqref{eq:G}, we have
	\begin{align}
		G &= [ 1/4(\alpha_4A\cdot F^2- \alpha_2^2 F\cdot DF+\alpha_2^2DF \cdot F - \alpha_4 F^2 \cdot A) \nonumber\\
		&\quad  +1/16 (\alpha_2 F\cdot A \cdot A \cdot A-  \alpha_2 A\cdot F \cdot A \cdot A+  \alpha_2 A\cdot A \cdot F \cdot A-\alpha_2 A\cdot A \cdot A \cdot F )\nonumber\\
		&\quad - 1/2 \alpha_4 DF^2 +1/8(\alpha_2 A\cdot A \cdot DF +\alpha_2 A\cdot DF \cdot A + \alpha_2 DF\cdot A \cdot A) \nonumber\\
		&\quad  +1/8(-\alpha_2^2 A \cdot F \cdot F + \alpha_2^2 F \cdot A \cdot F - \alpha_2^2F \cdot F \cdot A) \nonumber\\
		& \quad-1/32 A\cdot A \cdot A \cdot A \cdot A \nonumber \\
		&\quad +1/2\alpha_4 \diff F^2 +1/8( -\alpha_2 \diff F \cdot A \cdot A+\alpha_2 \diff A \cdot F \cdot A-\alpha_2\diff A \cdot A \cdot F )]\diff \omega \nonumber \\
		&=-1/32( 2A(\diff A)^2+(\diff A)A(\diff A)+2(\diff A)^2 A + 2(\diff A)A^3 \nonumber\\
		&\hspace{2cm}+ A(\diff A)A^2+ A^2(\diff A)A +2A^3(\diff A) + 2A^5 )\diff \omega.
	\end{align}
	Thus we obtain
	\begin{align}
		\lim_{a \to 0} a^6\sum_x \mathcal{A}_{\mathrm{gauge}}(x) &=\frac{i}{48\pi}\int \tr \Bigl(\frac{1}{20}(2(\diff A)^2A+(\diff A)A(\diff A)+2A(\diff A)^2) \nonumber \\
			&\quad+\frac{1}{20}(2(\diff A)A^3+A(\diff A)A^2+A^2(\diff A)A+2A^3(\diff A))+\frac{1}{10}A^5\Bigr)\diff \omega.
	\end{align}
	\end{enumerate}
	It turns out from the above results that the gauge anomalies in two, four, and six dimensions in the continuum limit obtained here are equivalent to those known in the continuum theory up to total derivatives (for a review of the gauge anomaly, see Ref.~\cite{AlvarezGaume:1983cs}).
	
	\section{Conclusion}
	In this paper, we generalized the result in Ref.~\cite{Makino:2017pbq}, in which the gauge anomaly of the four-dimensional effective theory is calculated; i.e., we performed the explicit calculation of the arbitrary even-dimensional gauge anomaly with the chiral overlap operator in the continuum limit.
	The resultant expressions in two, four, and six dimensions are found to be equivalent to those known in the continuum theory up to total derivatives.
	If the gauge field is evolved by the gradient flow, the total effective action is gauge invariant, and the anomalies are cancelled by the cross terms of the gauge fields $A$ and $A_{\star}$.
	This means that the parity-odd part of the cross terms corresponds to the Chern-Simons term. 
	Thus the parity-odd part of the cross terms vanishes if the anomaly cancellation condition is satisfied.
	
	\section*{Acknowledgements}
	The author would like to thank Takeo Moroi and Natsumi Nagata for helpful discussion and advice. 
	\bibliographystyle{JHEP}
	\bibliography{ref}
\end{document}